\documentclass{article}

\usepackage{PRIMEarxiv}

\usepackage[utf8]{inputenc} 
\usepackage[T1]{fontenc}    
\usepackage{hyperref}       
\usepackage{url}            
\usepackage{booktabs}       
\usepackage{amsfonts}       
\usepackage{nicefrac}       
\usepackage{microtype}      
\usepackage{lipsum}
\usepackage{fancyhdr}       
\usepackage{graphicx}       
\usepackage{amsmath}
\usepackage{algorithm}
\usepackage{algpseudocode}
\usepackage{float} 

\graphicspath{{media/}}     

\title{Template-Fitting Meets Deep Learning: Redshift Estimation Using Physics-Guided Neural Networks
}

\author{
  \href{https://orcid.org/0009-0005-6152-3520}{\includegraphics[scale=0.1]{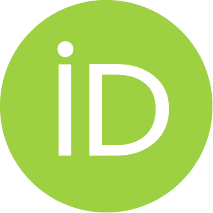}\hspace{1mm}
  Jonas Chris Ferrao} \\
  Department of Computer Engineering \\
  Don Bosco College of Engineering \\
  Fatorda, India \\
  \texttt{jonasferrao21@gmail.com} \\
  \AND
  \href{https://orcid.org/0009-0004-1447-4298}{\includegraphics[scale=0.1]{orcid.pdf}\hspace{1mm}
  Dickson Dias} \\
  Department of Computer Engineering \\
  Don Bosco College of Engineering \\
  Fatorda, India \\
  \texttt{diasdickson94@gmail.com} \\
  \And
  \href{https://orcid.org/0009-0001-1414-8490}{\includegraphics[scale=0.1]{orcid.pdf}\hspace{1mm}
  Pranav Naik} \\
  Department of Computer Engineering \\
  Don Bosco College of Engineering \\
  Fatorda, India \\
  \texttt{naikpranav110803@gmail.com} \\
  \And
  \href{https://orcid.org/0009-0003-5819-6843}{\includegraphics[scale=0.1]{orcid.pdf}\hspace{1mm}
  Glory D'Cruz} \\
  Department of Computer Engineering \\
  Don Bosco College of Engineering \\
  Fatorda, India \\
  \texttt{dcruzglory03@gmail.com} \\
  \And
  Anish Naik \\
  Department of Computer Engineering \\
  Don Bosco College of Engineering \\
  Fatorda, India \\
  \texttt{nanish909@gmail.com} \\
  \And
  Siya Khandeparkar \\
  Department of Computer Engineering \\
  Don Bosco College of Engineering \\
  Fatorda, India \\
  \texttt{siya.naik@dbcegoa.ac.in} \\
  \And
  Manisha Gokuldas Fal Dessai \\
  Department of Computer Engineering \\
  Don Bosco College of Engineering \\
  Fatorda, India \\
  \texttt{manisha.faldessai@dbcegoa.ac.in}
}

\begin{document}
\maketitle

\begin{abstract}
Accurately estimating photometric redshifts is crucial for observational cosmology, especially in large-scale surveys where spectroscopic redshifts are unfeasible. Template fitting and machine learning are the two primary methods for estimating photometric redshifts, each with its own set of advantages and drawbacks. This research proposes a novel approach that combines template fitting and deep learning techniques, leveraging physics-guided neural networks. The methodology involves integrating galaxy spectral energy distribution templates with neural network architectures to incorporate physical priors into the learning process. We developed a multimodal network that uses a cross-attention mechanism to integrate photometric and image data, and incorporated Bayesian layers to predict uncertainty. We used the publicly available PREML dataset, which comprises \(\sim \)400,000 galaxies from the Hyper Suprime-Cam survey's PDR3 release. This dataset includes 5-band photometry, multi-band images, and spectroscopic redshifts, which we use to evaluate the effectiveness of our approach. Our method achieves an RMS error of 0.0507, a $3\boldsymbol{\sigma}$ catastrophic outlier rate of 0.13\% and a bias of 0.0028. Our results demonstrate that our method meets 2 of the 3 LSST requirements for $z < 3$. These results demonstrate the effectiveness of combining template-fitting and data-driven approaches for photometric redshift estimation, which will be essential for future cosmological surveys.
\end{abstract}
\section{Introduction}
Redshift estimation is a fundamental task in modern astrophysics, enabling astronomers to determine the distances and velocities of celestial objects. Accurate redshift measurements are crucial for studying the large-scale structure of the universe \cite{LargeScaleStructure}, and galaxy evolution \cite{finkelstein2015evolution}. While spectroscopic redshift estimation remains the gold standard because of its high precision, it is not feasible at scale due to its resource-intensive nature. As next-generation surveys like LSST \cite{ivezic2019lsst}, and Euclid \cite{Euclid} are set to produce petabytes of data, accurate and efficient alternatives have become essential.

Photometric redshift (photo-z) estimation methods offer a promising solution by using broadband photometry to infer redshifts. There are 2 main methods for photometric redshift estimation:

- Machine Learning: In this approach, Machine Learning algorithms are trained to predict spectroscopic redshift from observed photometric data. These algorithms learn to map photometric inputs to their corresponding spectroscopic redshifts. These methods are highly accurate and fast, especially with the advent of high-performance computing such as GPUs and TPUs. However, the main requirement is to have a training set that is representative of the observed data. Hence, most ML algorithms fall short for higher redshifts where limited training data are available.

- Template Fitting: This method works by matching spectral energy distribution templates of galaxies to observed photometric data to obtain redshift estimates. The method is grounded in physical principles and does not require a training set, which makes it applicable across a broad redshift range. However, in practice, SED-based approaches struggle due to factors such as template selection, corrections for extinction and reddening, and spectral evolution assumptions, as well as other potential sources of error.

The LSST Science Requirements Document \cite{ivezic2018lsst1} specifies the following targets for photometric redshift (photo-z) estimation in a magnitude-limited sample of galaxies with i < 25 over the redshift range 0.3 < z < 3.0: 

1:- RMS error < 0.2

2:- Fraction of catastrophic outliers < 10\%,  

3:- Bias < 0.003

Currently, no approach satisfies all these requirements for z < 3. In addition to these requirements, methods to reject outliers must also be developed.

There exist numerous approaches for redshift estimation, among these Convolutional neural networks (CNNs) have become a cornerstone of photo-z estimation due to their ability to extract morphological features from galaxy images. \cite{Pasquet2019} developed a deep CNN to estimate photo-z for galaxies in the SDSS Main Galaxy Sample at z < 0.4 using ugriz images. Similarly, \cite{Schuldt2021} introduced NetZ, another CNN based method optimized for high redshifts (z > 2), leveraging morphological information to outperform traditional methods where spectroscopic data is scarce. \cite{Yao2023} proposed QPreNet, a novel network that integrates images and photometric data to estimate quasar redshifts. Their fused-feature approach outperformed single-modality methods, demonstrating the power of combining data types. \cite{AitOuahmed2023} further advanced CNN-based methods with a multimodal approach that processes subsets of photometric bands in parallel, achieving statistically significant improvements in precision across multiple datasets.

Bayesian neural networks (BNNs) offer robust uncertainty estimates, making them ideal for photo-z applications where reliability is critical. \cite{Jones2024b} presented a BNN model for the LSST, achieving accurate predictions with well-calibrated uncertainties in the redshift range 0.3 < z < 1.5. Their model met two of three LSST photo-z requirements and outperformed non-Bayesian alternatives. \cite{Jones2024a} extended this work with a Bayesian CNN (BCNN), which met the LSST requirements up to z < 1.5 by leveraging galaxy images, outperforming non-image-based methods in scatter and outlier rates. \cite{Zhou2022} applied BNNs to the China Space Station Telescope (CSST) survey, demonstrating that CNN-based BNNs extract more information from images than multilayer perceptrons (MLPs) using flux data alone. Their hybrid network, which combined flux and images, further reduced outlier fractions and improved prediction confidence.

Long short-term memory (LSTM) networks have shown promise in capturing complex relationships in photometric data. \cite{Luo2024} proposed an LSTM-based model for the CSST survey, outperforming template-fitting methods such as EAZY and other ML approaches such as weighted random forest (WRF). Their model automatically learns relationships across wavelengths and generates probability density functions (PDFs) using Monte Carlo dropout, enhancing confidence in predictions. \cite{Dey2021} introduced a deep capsule network for SDSS images, jointly predicting photo-z and morphological types with fewer parameters than traditional methods. Their model achieved comparable or better accuracy and provided interpretable feature encodings, demonstrating the potential of compact architectures.

Multimodal methods have consistently outperformed single-modality approaches. \cite{Hong2022} developed PhotoRedshift-MML, a multimodal ML method for quasar photo-z estimation, achieving an accuracy of 84.45\% for |z| < 0.1 and reducing the RMS error compared to photometric-only methods. \cite{Henghes2022} explored mixed-input CNNs, finding that their CNN inception module outperformed traditional ML algorithms, particularly for galaxies with z < 0.3, though at higher computational cost. \cite{Hoyle2016} proposed a deep neural network (DNN) approach using full galaxy images, achieving performance comparable to AdaBoost but with a slightly higher outlier rate. \cite{DIsanto2017} advanced this with a deep convolutional mixture density network (DCMDN), predicting PDFs directly from multi-band imaging without pre-classification. Their model outperformed feature-based models in both traditional and probabilistic metrics.

In this study, we propose a novel multimodal approach for highly accurate photometric redshift estimation. Our model incorporates Bayesian layers for uncertainty quantification, convolutional layers for image-based inputs, and a cross-attention mechanism for optimised multimodal integration and prediction.

We have 3 goals in this work.

1) Develop an optimised neural architecture that delivers state-of-the-art redshift predictions.

2) Integrate physics-based SED fitting within a deep-learning pipeline.

3) Establish a new benchmark against LSST requirements up to z = 3.

Section 2 describes the dataset used in this study. Section 3 details the architecture and training procedure of the neural network model, including the integration of SED-based loss functions. This section contains technical content and may be skipped by readers primarily interested in results. Section 4 presents the photometric redshift estimation results on the test data and compares them with other models. Finally, Section 5 summarises the findings and outlines future directions.

\section{Data}
\label{sec:data}

This study leverages data from the Hyper Suprime-Cam Subaru Strategic Program (HSC-SSP) Public Data Release 3 (PDR3) \cite{aihara2022third}, which provides high-quality imaging and photometry in five broad-band filters (g, r, i, z, y). Designed to simulate the depth and data characteristics of upcoming large-scale surveys like the Vera C. Rubin Observatory's Legacy Survey of Space and Time (LSST), PDR3's Wide layer provides complete coverage over 670 square degrees. It achieves a median seeing of ~0.6 arcseconds in the i-band and reaches 5$\sigma$ point-source detection limits of approximately 26 magnitudes. This dataset is particularly suitable for developing and benchmarking photometric redshift (photo-z) prediction models due to its depth and photometric accuracy, which closely align with LSST goals.

\subsection{Photometric and Spectroscopic Data}
\label{subsec:phot_spec_data}

Photometric data are accessed through the HSC PDR3 database via the HSC CasJobs interface (\url{https://hsc-release.mtk.nao.ac.jp/datasearch/}). We retrieve cModel magnitudes, fluxes, and their associated errors for the five optical bands (g, r, i, z, y), corresponding to the database keys:

\begin{itemize}
    \item Magnitudes: \texttt{\{grizy\}\_cmodel\_mag},
    \item Magnitude Errors: \texttt{\{grizy\}\_cmodel\_magerr}
    \item Fluxes: \texttt{\{grizy\}\_cmodel\_flux}
    \item Flux errors: \texttt{\{grizy\}\_cmodel\_fluxerr}
    \item Spectroscopic Redshift: \texttt{specz\_redshift}
    \item Spectroscopic Redshift Error: \texttt{specz\_redshift\_err}
\end{itemize}

To ensure data quality, we apply selection criteria similar to those used in GalaxyML Dataset \cite{do2024galaxiesmldatasetgalaxyimages}. We require detections in all five bands and apply quality cuts to remove unreliable photometric measurements and poorly measured spectroscopic redshifts.


These cuts ensure that the dataset is free from outlier photometric measurements, saturated pixels, and unreliable spectroscopic redshifts, resulting in a clean and robust sample for analysis.

\subsection{Image Data}
\label{subsec:image_data}

\begin{figure}[H]
\centering
\includegraphics[width=\textwidth]{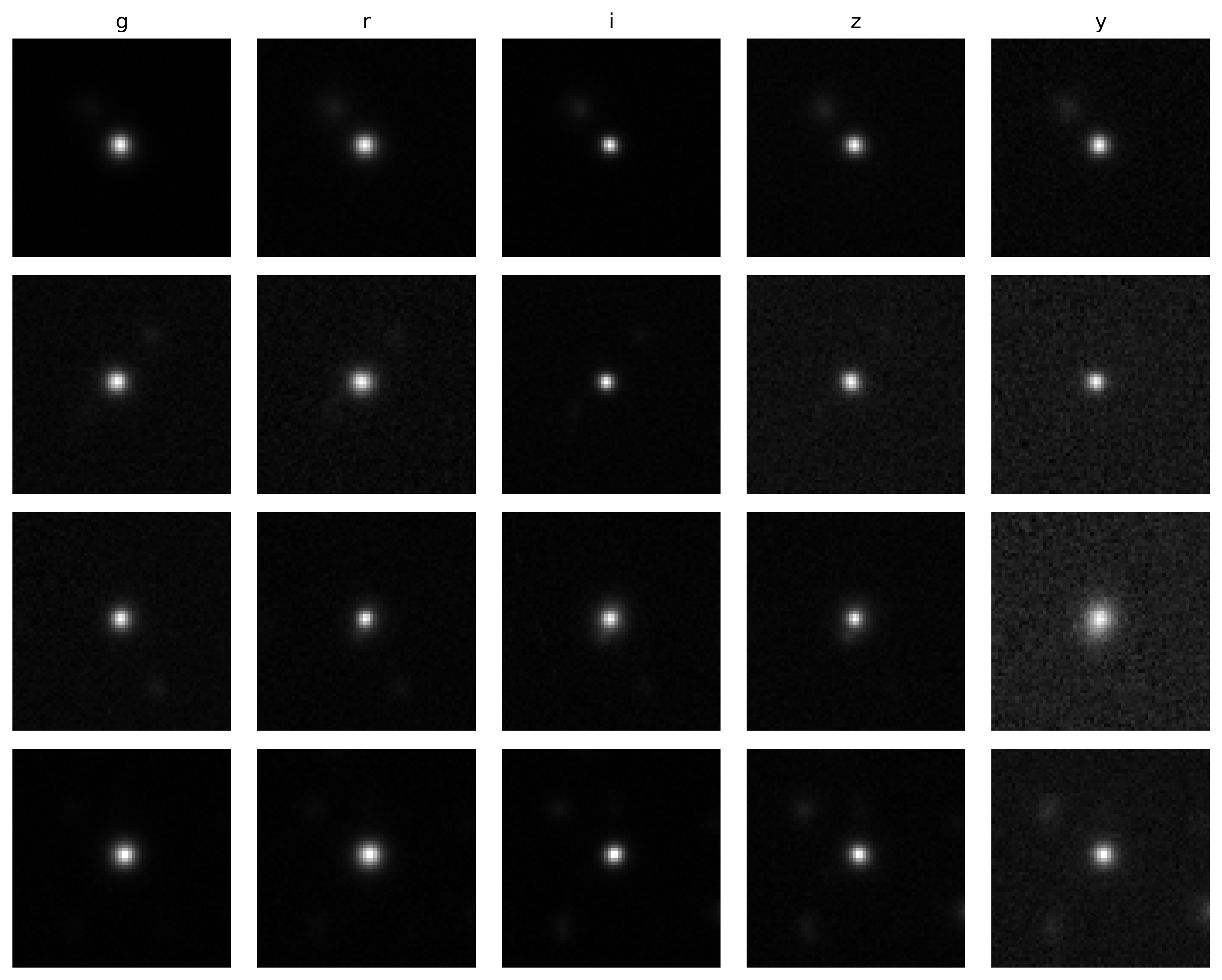}
\caption{Examples of HSC galaxy cutouts across the five photometric bands (g, r, i, z, y) for two galaxies at high redshift (top) and two at low redshift (bottom), specifically at $z = 2.441$, $z = 2.614$, $z = 0.772$, and $z = 0.815$ (ordered top to bottom). The visual similarity between galaxies at widely different redshifts highlights the difficulty of estimating redshift purely from imaging data.}
\label{fig:galaxy_redshift_comparison}
\end{figure}

The image data are obtained using the HSC PDR3 cutout service (\url{https://hsc-release.mtk.nao.ac.jp/das_cutout/pdr3/}). For each galaxy in the photometric catalog, we generate image cutouts centered on the right ascension (RA) and declination (Dec) coordinates, using a coordinate list derived from the photometric entries. The cutouts are retrieved with a size of $sw = 0.0896$ deg and $sh = 0.0896$ deg, corresponding to 64$\times$64 pixel images with a pixel scale of 0.168 arcsec/pixel, in the \texttt{pdr3\_wide} rerun. Images are collected for all five bands (g, r, i, z, y) and saved as FITS files.
Figure~\ref{fig:galaxy_redshift_comparison} shows examples of HSC galaxy images in all five photometric bands (g, r, i, z, y) for four galaxies at different redshifts. The top two rows display high-redshift galaxies at $z = 2.441$ and $z = 2.614$, while the bottom two rows show low-redshift galaxies at $z = 0.772$ and $z = 0.815$. Despite the significant difference in redshift, these galaxies can appear remarkably similar in their morphology and appearance across the different bands, highlighting the challenge in photometric redshift estimation.

\subsection{Final Dataset}
\label{subsec:final_dataset}

\begin{figure}[H]
    \centering
    \includegraphics[width=0.5\textwidth]{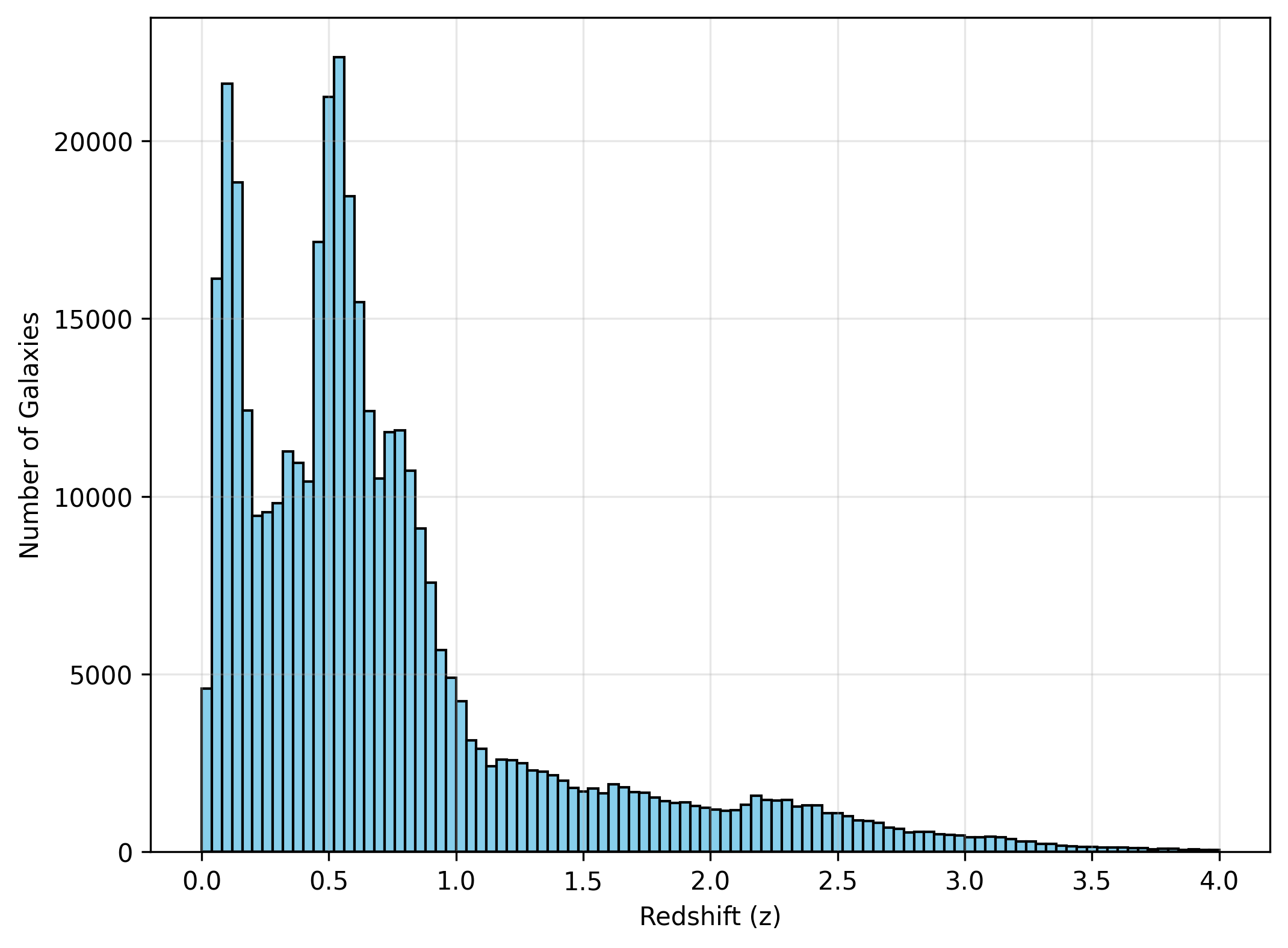}
    \caption{The N(z) distribution of the spectroscopic redshift sample from the HSC PDR3, comprising 395,585 galaxies. The distribution shows peaks at $z \approx 0.1$ and $z \approx 0.6$, with galaxies ranging from $0.01 < z < 4$.}
    \label{fig:nz_distribution}
\end{figure}

After applying the quality cuts and image processing steps, the final dataset consists of 395,585 galaxies with five-band grizy photometry, corresponding 64$\times$64 pixel images, and spectroscopic redshifts. The redshift distribution, shown in Figure~\ref{fig:nz_distribution}, spans $0.01 < z < 4$, with the majority of galaxies lying between $z = 0.01$ and $z = 2.5$, peaking at $z \approx 0.1$ and $z \approx 0.6$. The dataset is divided into 70\% for training (276,910 galaxies), 10\% for validation (39,558 galaxies), and 20\% for testing (79,117 galaxies).

The photometric measurements, image cutouts, and spectroscopic redshifts used for training, validation, and testing are all made publicly available as part of the PREML dataset on Zenodo at \url{https://doi.org/10.5281/zenodo.15426393}. This ensures full reproducibility of the experiments conducted in this study. The original photometric data can also be accessed through the HSC PDR3 database.

\section{Methodology}
We propose a multimodal neural network that jointly processes image cutouts and photometric magnitudes to estimate redshift. The architecture is composed of four main modules: the image block, the magnitude block, the attention block, and the prediction block. The model predicts the photometric redshift, its uncertainty, and the spectral energy distribution (SED) class.

\subsection{Neural Network Architecture}

\subsubsection{Magnitude Block} This block processes the photometric magnitudes using a stack of fully connected layers interleaved with PReLU activations and dropout layers. Dropout randomly deactivates neurons during training to prevent overfitting. PReLU allows learning the activation slope, improving upon standard ReLU.

\subsubsection{Image Block} The image block consists of several 2D convolutional layers, each followed by ReLU activation and max pooling. It extracts hierarchical spatial features from the image cutouts.

\subsubsection{Attention Block} We use cross-attention to allow magnitude embeddings to attend to the image features. This enables the model to identify which image regions are most relevant to each magnitude. The attention mechanism is implemented using multi-head attention with 8 heads and dropout. The attended image features are flattened and concatenated with the magnitude features.

\subsubsection{Prediction Block} The combined features are passed to three separate heads:

Mean Regression Head: A Multilayer Perceptron network with ReLU and dropout, ending in a ReLU activation to ensure non-negative redshift predictions.

Variance Estimation Head: A Bayesian neural network with three BayesLinear layers. All layers have priors with $\mu = 0$ and $\sigma = 0.01$, except the final layer that uses $\mu = 0.8$ and $\sigma = 0.1$.

Classification Head: A dense layer outputs softmax probabilities over SED template classes.

\begin{figure}[H]
    \centering
    \includegraphics[width=0.8\textwidth]{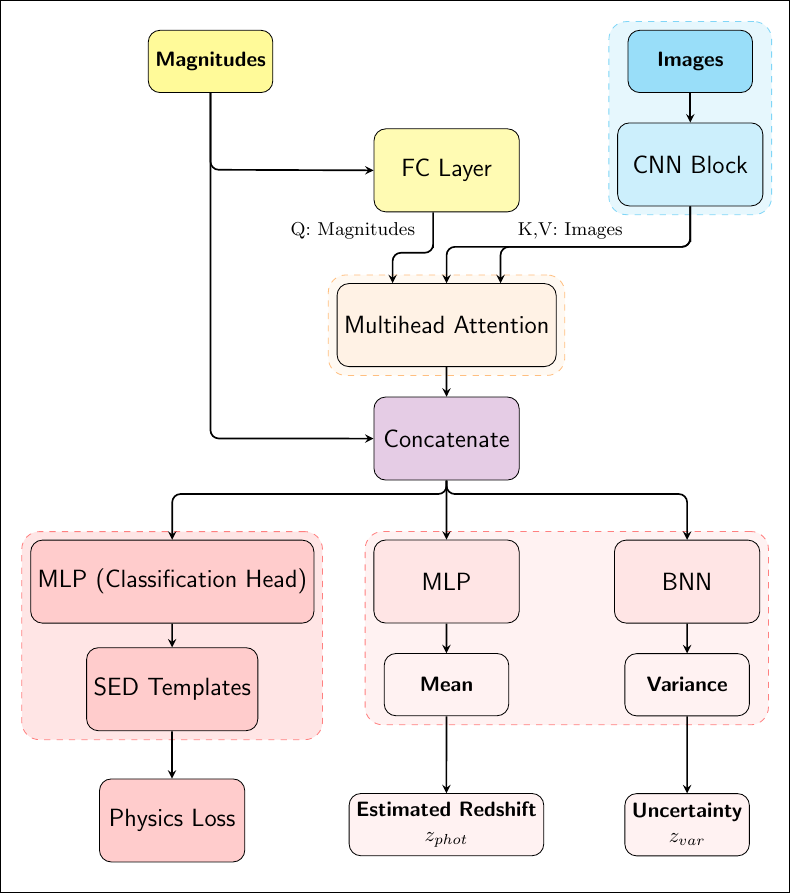}
    \caption{The architecture of the proposed Physics-Guided Neural Network}
    \label{fig:Architecture}
\end{figure}

\newpage
\subsection{Template Fitting Integration}
Spectral Energy Distribution template fitting involves matching the observed fluxes of a galaxy to those predicted by a set of redshifted SED models. These SED templates are typically derived from stellar population synthesis models or empirical observations of galaxies. The templates are redshifted across a predefined range and convolved with the transmission curves of the survey filters to simulate the observed photometry. Finally, a redshift estimate is obtained by identifying the template and redshift combination that best fits the observed data, often using a $\chi^2$ minimization technique.

\begin{figure}[H]
    \centering
    \includegraphics[width=1\textwidth]{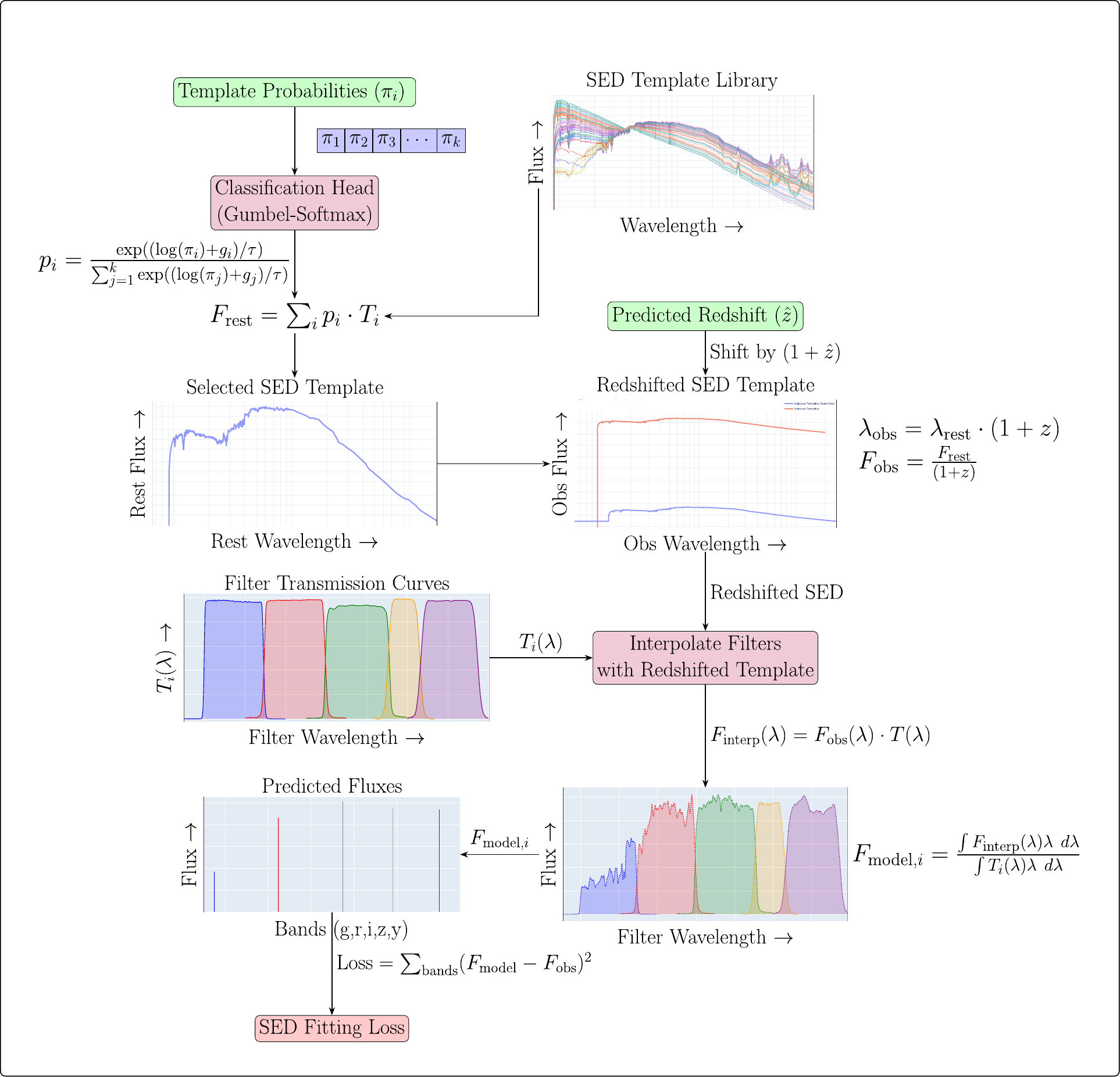}
    \caption{The SED Template Fitting Process}
    \label{fig:SED}
\end{figure}
Template Combination: The network predicts probabilities for a set of spectral templates. These probabilities are used to create a weighted combination of template fluxes, representing the model's estimate of the galaxy's spectral energy distribution in the rest frame.

Redshifting: The combined template is then redshifted on the basis of the network's predicted redshift. This involves calculating the luminosity distance and adjusting the fluxes to simulate the observed fluxes at that redshift.

Filter Integration: The redshifted SED is integrated over the filter response curves to predict the flux in each observed band. This involves interpolating the SED to match the filter wavelengths and using a trapezoidal rule for integration.

Magnitude Calculation and MSE Loss: The predicted fluxes are converted to magnitudes, and the mean squared error between the predicted magnitudes and the observed magnitudes is computed. This MSE loss penalizes discrepancies between the model's physics-based flux predictions and the actual observations. The code includes a scaling factor $a$ that minimizes the difference between the predicted and target fluxes. Clamping and adding a small value (1e-32) prevent numerical instability from zero or negative flux values.

\begin{algorithm}
\caption{Physics-Guided Loss Calculation}
\begin{algorithmic}[1]
\Require $z_{\text{pred}}$, $F_{\text{obs}}$, $\sigma$, $\{\mathcal{T}_i(\lambda)\}$, $P_{\text{soft}}$, $\{F^{\text{rest}}_j(\lambda)\}$, $\lambda^{\text{rest}}$, $m_{\text{obs}}$
\Ensure $\text{SED}_{\text{loss}}$
\State $F^{}_{\text{rest}}(\lambda) \gets \text{einsum}(P_{\text{soft}}, \{F^{\text{rest}}_j(\lambda)\})$ \Comment{Soft combination of rest-frame templates}
\State $\lambda_{\text{obs}} \gets \lambda^{\text{rest}} \cdot (1 + z_{\text{pred}})$
\State $F^{\text{obs}}_{\text{model}}(\lambda) \gets \dfrac{F^{\text{rest}}_{\text{model}}(\lambda)}{(1 + z_{\text{pred}})}$ \Comment{Apply redshift and dimming}
\For{each band $i$}
    \State Interpolate $F^{\text{obs}}_{\text{model}}(\lambda)$ to $\mathcal{T}_i(\lambda)$
    \State $F_{\text{model},i} \gets \dfrac{\int F^{\text{obs}}_{\text{model}}(\lambda) \mathcal{T}_i(\lambda) \lambda \, d\lambda}{\int \mathcal{T}_i(\lambda) \lambda \ d\lambda}$ \Comment{Synthetic flux}
\EndFor
\State $s \gets \dfrac{\sum_i \left( \dfrac{F_{\text{obs},i} \cdot F_{\text{model},i}}{\sigma_i^2} \right)}{\sum_i \left( \dfrac{F_{\text{model},i}^2}{\sigma_i^2} \right)}$ \Comment{Best-fit scaling factor}
\State $m_{\text{model},i} \gets -2.5 \log_{10}(s \cdot F_{\text{model},i}) + \text{const}$ \Comment{Convert to magnitudes}
\State $\text{SED}_{\text{loss}} \gets \text{MSE}(m_{\text{model}}, m_{\text{obs}})$
\State \Return $\text{SED}_{\text{loss}}$
\end{algorithmic}
\end{algorithm}

\subsection{Training Details}
The model was developed using the PyTorch and torchbnn packages. We trained the model on an RTX A6000 GPU with 48GB VRAM and an Intel Core i9 14900k with 192GB RAM. We trained for 30 epochs and observed rapid convergence, using the Nadam optimizer instead of Adam. We saved the model at every epoch and used the model with the lowest validation loss. A learning rate of 0.001 was used, and the batch size was 128. For the losses, we used a combination of multiple losses: RMSE, NLL, and SED losses.

\begin{equation}
\text{Total Loss} = \text{RMSE} + \text{Gaussian NLL} + \text{SED Loss} \cdot 0.1
\end{equation}

\subsection*{Individual Loss Components}

\begin{enumerate}
    \item \textbf{RMSE (Root Mean Squared Error)}
    \begin{equation}
    \text{RMSE} = \sqrt{\frac{1}{N} \sum_{i=1}^{N} (y_i - \hat{y}_i)^2}
    \end{equation}
    where \( y_i \) is the actual value, \( \hat{y}_i \) is the predicted value, and \( N \) is the number of samples.

    \item \textbf{Gaussian NLL (Negative Log-Likelihood)}
    \begin{equation}
    \text{Gaussian NLL} = \frac{1}{N} \sum_{i=1}^{N} \left[ \frac{(y_i - \mu_i)^2}{2\sigma_i^2} + \frac{1}{2} \ln (2\pi \sigma_i^2) \right]
    \end{equation}
    where \( \mu_i \) is the predicted mean, \( \sigma_i^2 \) is the predicted variance, and \( y_i \) is the actual value.

    \item \textbf{SED Loss}
    \begin{equation}
    \text{SED Loss} = \frac{1}{N} \sum_{i=1}^{N} (m_i - \hat{m}_i)^2
    \end{equation}
    where \( m_i \) is the actual magnitude, \( \hat{m}_i \) is the template predicted magnitude, and \( N \) is the number of samples.
\end{enumerate}

Table \ref{tab:hyperparameter_ranges} shows the set of hyperparameters along with the search space.

\begin{table}
  \caption{Hyperparameter details}
  \centering
  \begin{tabular}{lll}
    \toprule
    \textbf{Hyperparameter} & \textbf{Final Value} & \textbf{Tested Range} \\
    \midrule
    Optimizer & NAdam & Adam, NAdam,  \\
    Dropout & 0.5 & 0 - 0.5\\
    Learning rate & 0.001 & 0.01, 0.001, 0.0001 \\
    Loss function & Negative log likelihood & NLL, MSE, RMSE,  \\
    Activation functions & ReLU, PReLU & ReLU, PReLU, Softplus \\
    Number of epochs & 30 & 10–100 \\
    Batch sizes & 128 & 64, 128, 512, 1024 \\
    Number of Bayesian layers & 2 & 1–4 \\
    Gaussian initialiser & Mean: 0, 0.8, Standard Deviation: 0.01, 0.1 & Mean: 0-1, Standard Deviation: 0.001–0.1 \\
    \bottomrule
  \end{tabular}
  \label{tab:hyperparameter_ranges}
\end{table}

\section{Results}
To evaluate the performance of our Physics-Guided Neural Network (PGNN) model for photometric redshift estimation, we visualise its predictive performance using various statistical metrics. These metrics quantify the accuracy, robustness, and consistency of the model respectively, across the redshift spectrum.
\begin{table*}[h]
\centering
\caption{Metrics Used to Assess Model Performance}
\begin{tabular}{p{0.45\linewidth} p{0.45\linewidth}}
\toprule
\textbf{Point Metrics} & \textbf{Probabilistic Metrics} \\
\midrule

\textbf{RMS Error} & \textbf{PIT} \\
\begin{equation}
\sqrt{\frac{1}{n_{\text{gals}}} \sum_{\text{gals}} \left( \frac{z_{\text{phot}} - z_{\text{spec}}}{1 + z_{\text{spec}}} \right)^2}
\end{equation}
& 
\begin{equation}
\int_{-\infty}^{z_{\text{spec}}} p(z)\,dz
\end{equation}
\\[2ex]

\textbf{3$\sigma$ Catastrophic Outlier} & \textbf{Coverage} \\
\begin{equation}
O_c: |z_{\text{phot}} - z_{\text{spec}}| > 3\sigma_z
\end{equation}
& 
\begin{equation}
\frac{1}{n_{\text{gals}}} \sum_i (\bar{z}_{\text{pdf},i} - z_{\text{spec},i} < \sigma_{z,i})
\end{equation}
\\[2ex]

\textbf{Bias} & \\
\begin{equation}
b = \frac{z_{\text{phot}} - z_{\text{spec}}}{1 + z_{\text{spec}}}
\end{equation}
& \\
\bottomrule
\end{tabular}
\label{tab:model_metrics}
\end{table*}

\begin{figure}[H]
    \centering
    \includegraphics[height=4in]{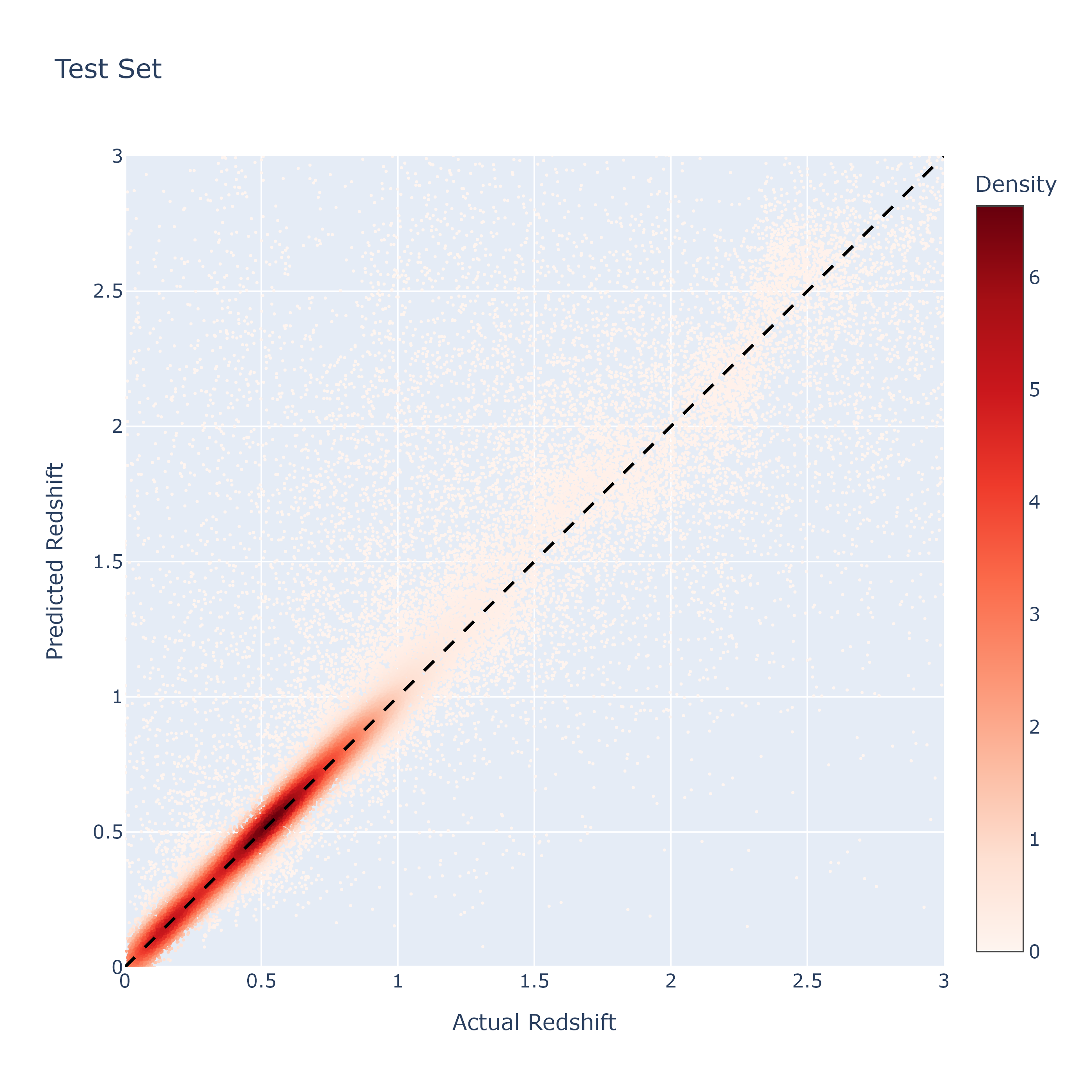}
    \caption{Scatter Plot of predicted vs actual redshifts} 
    \label{fig:fig1}
\end{figure}

Figure~\ref{fig:fig1} shows a scatter density plot that compares the predicted redshifts to the actual redshifts for the test set. The majority of predictions lie close to the $1:1$ diagonal line, indicating strong agreement between predicted and true values. The density is highest in the low-redshift range ($z < 1$), where the model performs well.

Deviations from the diagonal become more apparent at higher redshifts, with a visible spread above $z > 1.5$. This suggests increased uncertainty or reduced precision in that regime. There are some outliers beyond $z > 2.5$, but they are sparse and do not dominate the distribution.

\begin{figure}[H]
    \centering
    \includegraphics[height=3in]{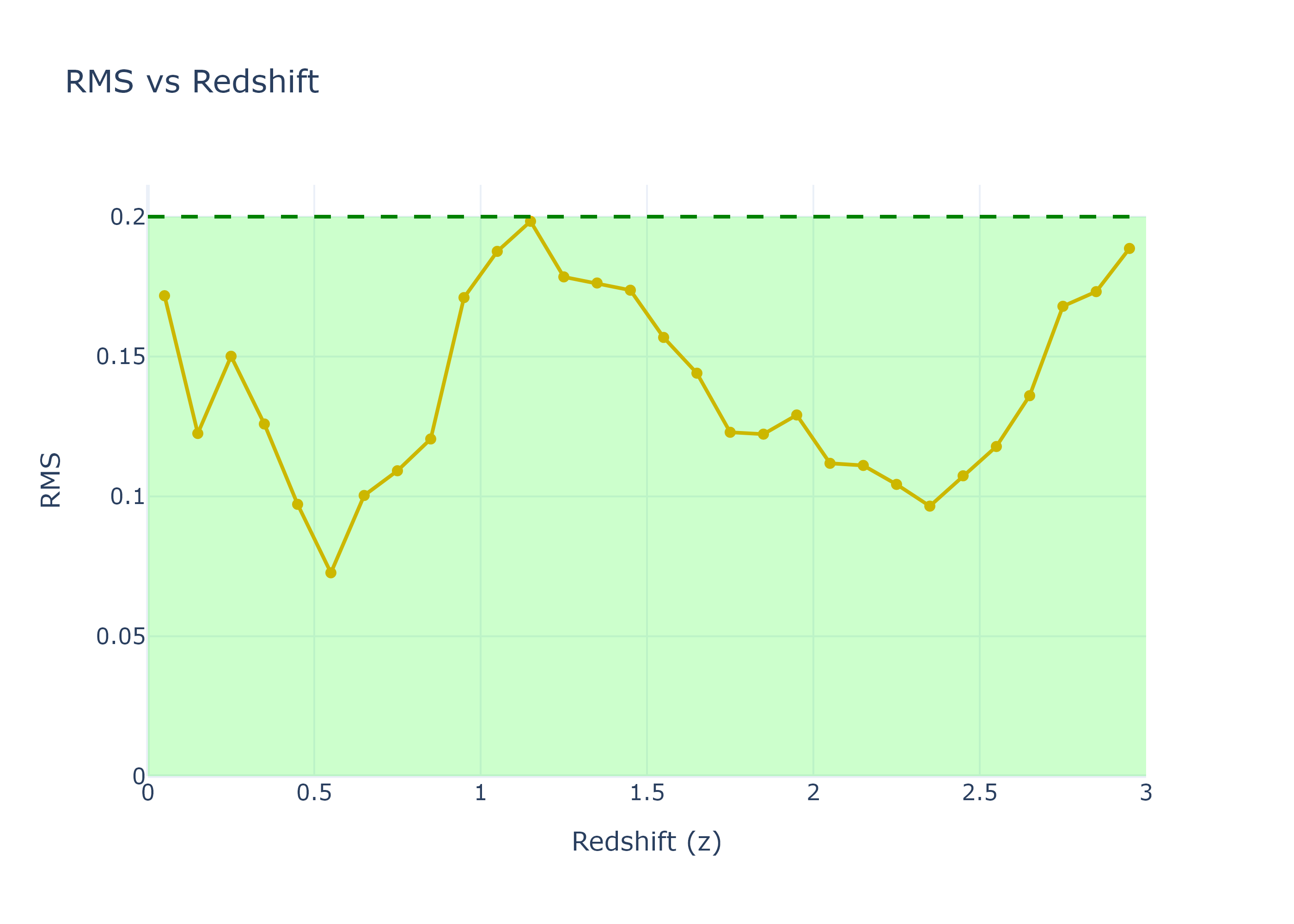}
    \caption{RMS}
    \label{fig:fig2}
\end{figure}

Figure~\ref{fig:fig2} presents the RMS error in varying redshift bins. The RMS values exhibit variation across the redshift range, remaining entirely below the LSST threshold of 0.2 (indicated by the green dashed line). The best performance is observed around $z \approx 0.5$ and $z \approx 2.4$, with RMS reaching values below 0.1. These results demonstrate that the model satisfies the LSST photometric redshift RMS quality criteria up to $z = 3$.
The elevated RMS around $z = 1.0$ may be attributed to degeneracies in the photometric colour space at that redshift, while the increased error at the boundaries is likely due to sparse data.

\begin{figure}[H]
    \centering
    \includegraphics[height=3in]{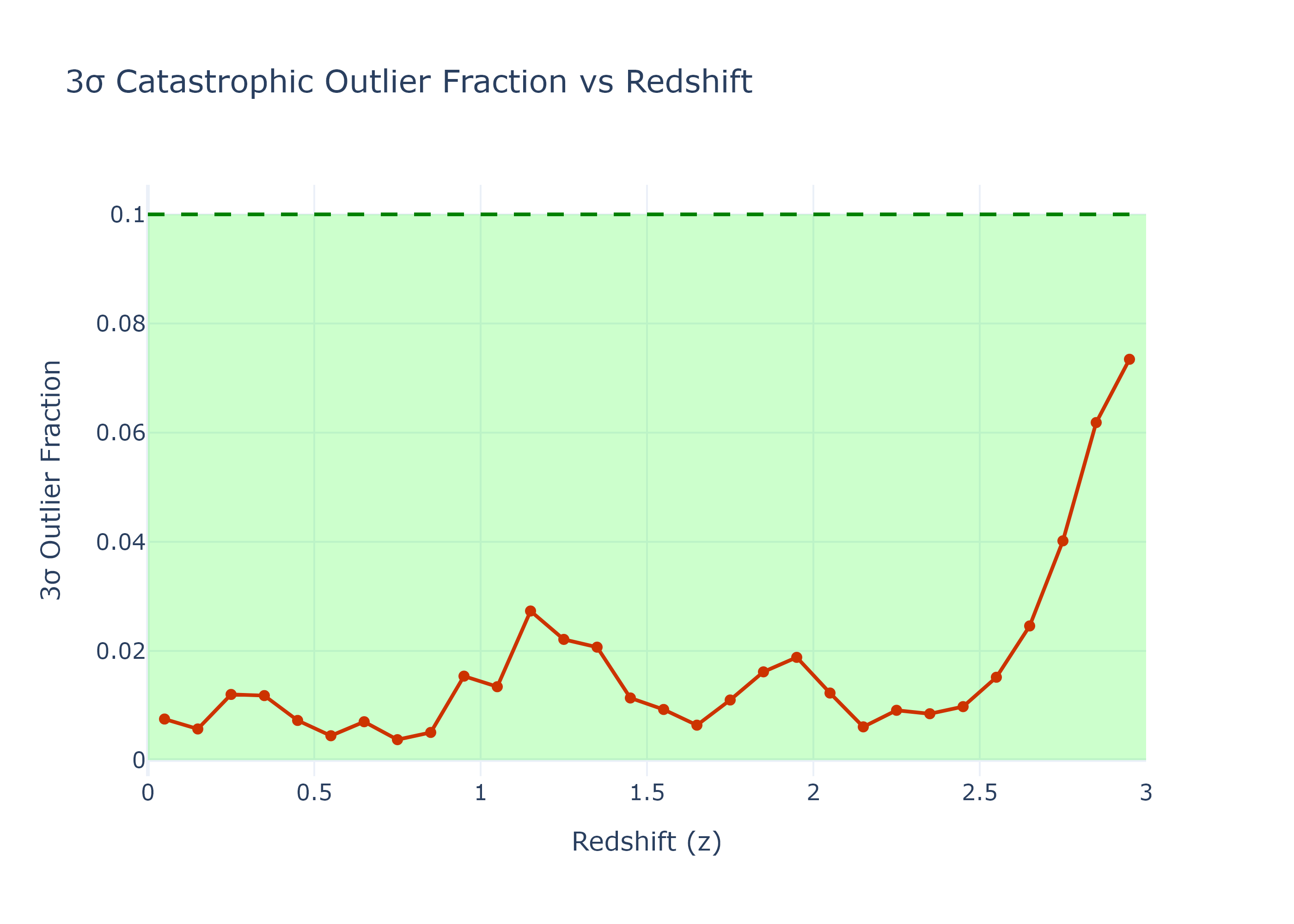}
    \caption{$3 \sigma$ Catastrophic Outliers}
    \label{fig:fig3}
\end{figure}

Figure~\ref{fig:fig3} presents the 3$\sigma$ catastrophic outlier fraction as a function of redshift. The model remains well below the LSST limit of 0.1 throughout the redshift range $0 < z < 3$, with outlier fractions typically under 0.02. A gradual rise is observed after $z = 2.6$, peaking at approximately 0.075 near $z \approx 2.95$, likely due to limited data in that region  The low outlier rate across the board suggests strong model generalisation with minimal failure cases at higher redshifts.

\begin{figure}[H]
    \centering
    \includegraphics[height=3in]{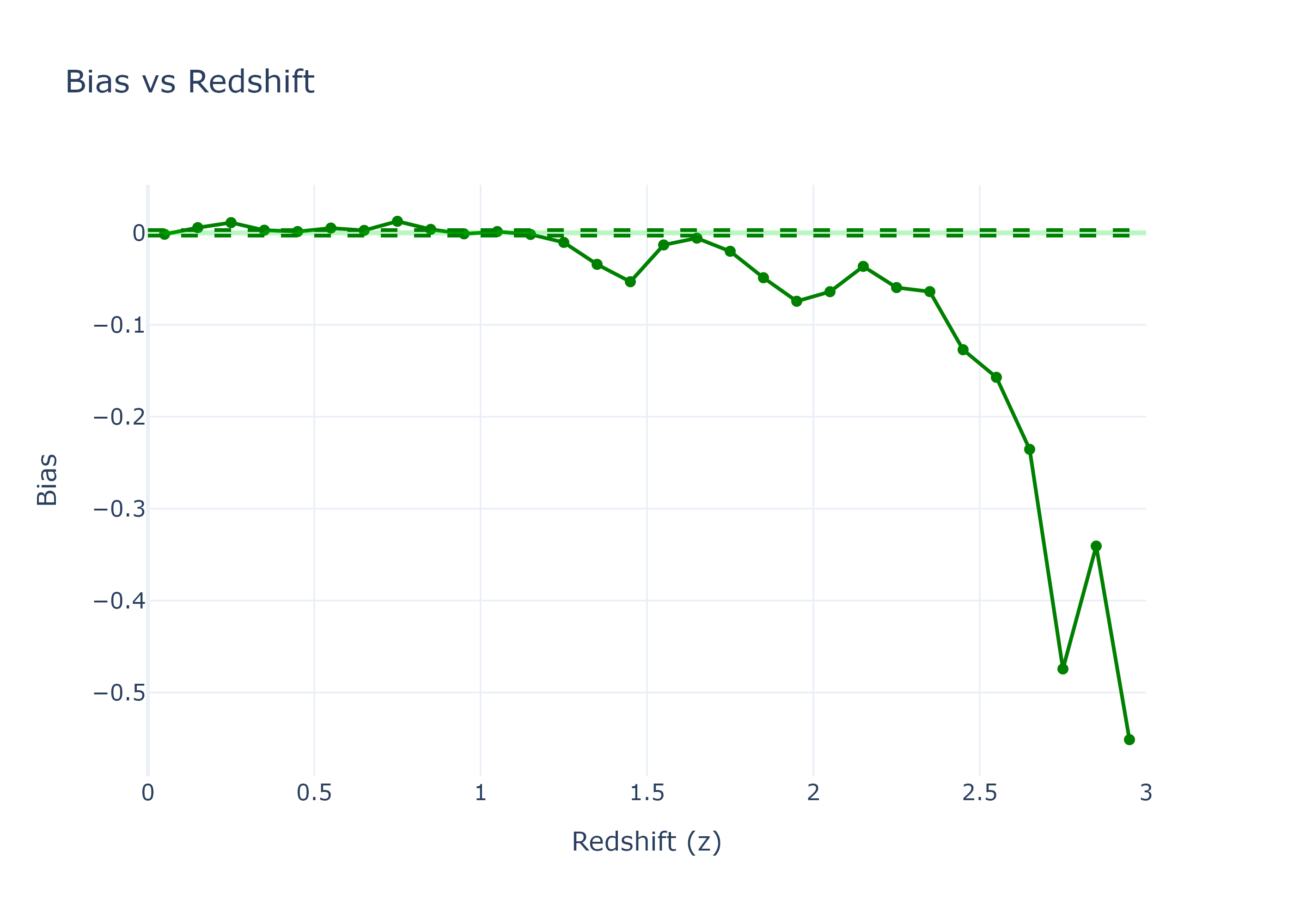}
    \caption{Bias}
    \label{fig:fig4}
\end{figure}

Figure~\ref{fig:fig4} shows the bias in redshift estimation across the redshift range. The bias remains close to zero up to $z \approx 2$, indicating accurate and unbiased predictions in the lower and intermediate redshift regimes. Beyond $z = 2$, the bias begins to drift negatively, reaching values below $-0.5$ near $z \approx 2.95$. This trend suggests a consistent underestimation of redshift at higher values, which may be attributed to reduced training data density or photometric degeneracies in that range.

\begin{figure}[H]
    \centering
    \includegraphics[height=2.3in]{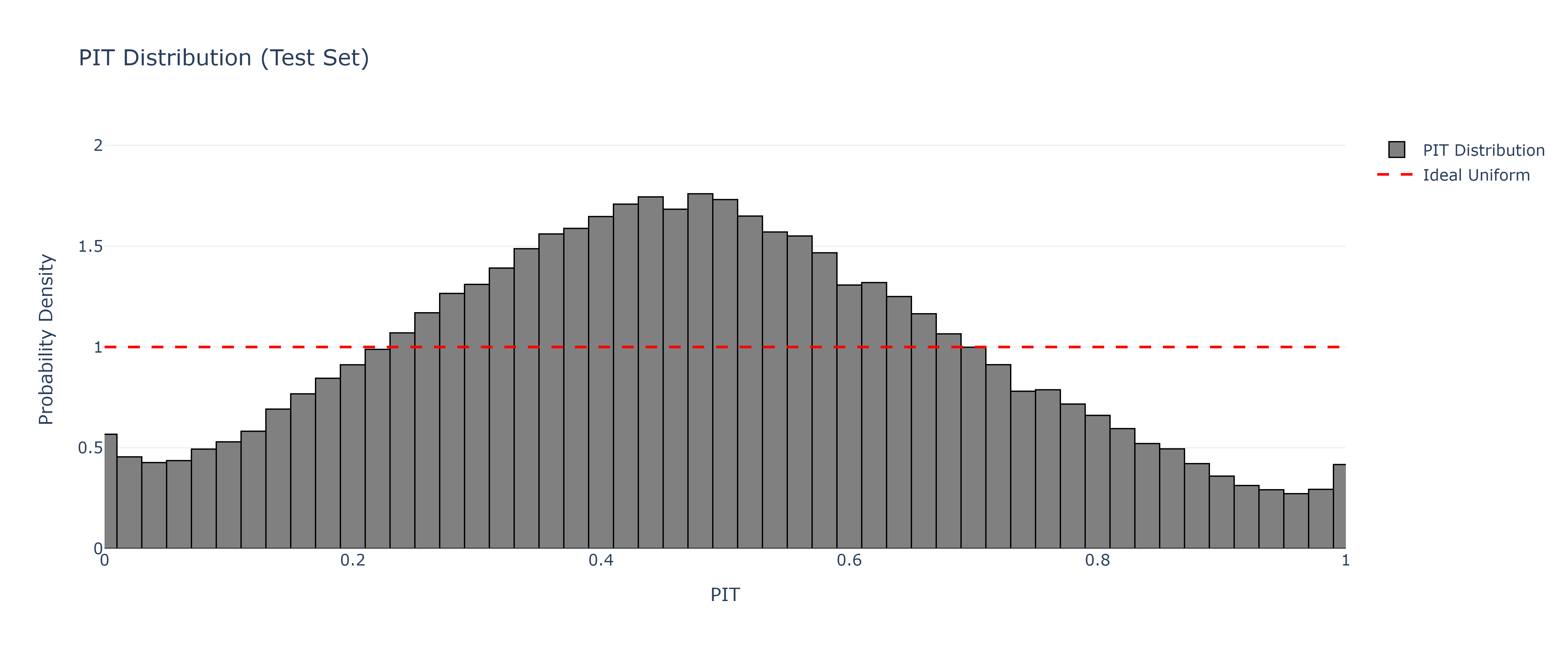}
    \caption{PIT}
    \label{fig:fig5}
\end{figure}

Figure~\ref{fig:fig5} presents the Probability Integral Transform (PIT) distribution for the test set. Ideally, a well-calibrated model should produce a PIT distribution that approximates a uniform distribution, as shown by the red dashed line. The observed distribution is centred with a slight taper at both ends, which indicates that the model is generally well-calibrated, though it may slightly underestimate the true uncertainty range in some cases. Overall, this suggests the model is effective in capturing the predictive uncertainties.


\section{Conclusion}
In this research, we presented a novel framework for photometric redshift estimation, combining Spectral Energy Distribution (SED) template fitting with deep learning. Our hybrid approach addresses traditional template-fitting methods limited adaptability to diverse galaxy types, and dependence on predefined templates. By integrating a multimodal deep neural network trained on enhanced high-redshift data, the model learns complex, non-linear mappings while retaining the interpretability and physical priors of SED models. Experiments conducted on benchmark datasets demonstrate improved redshift accuracy, reduced outlier rates, and enhanced generalisability across redshift ranges up to z = 3. This framework offers a scalable and robust solution for upcoming large-scale surveys like LSST and Euclid. While our approach shows strong performance for low-redshift galaxies, it exhibits a systematic bias at higher redshifts (z > 1.5). This bias can be mitigated with expanded and more representative training datasets, particularly from deep spectroscopic surveys that cover a broader range of galaxy types and redshifts. The SED templates used in this study were effective for low-redshift populations, but extending the model to higher redshifts will require incorporating high-redshift-specific templates that better capture the diversity of extreme galaxy types. Further improvements could involve embedding additional physical constraints, such as stellar mass–metallicity relations. We released our framework as an open-source tool to promote collaborative development. Ultimately, these enhancements aim to establish hybrid approaches as standard practice in observational cosmology, enabling more accurate redshift estimation and broader application of physics-informed machine learning methods in astrophysics.

\section*{Acknowledgments}
This work was supported by our institution's computational infrastructure, including access to an RTX A6000 GPU. This research makes use of photometric and spectroscopic data from the Hyper Suprime-Cam Subaru Strategic Program (HSC-SSP), specifically the Public Data Release 3 (PDR3). This research did not receive any external funding.

\section*{Data and Code Availability}
The original photometric and spectroscopic data used in this study were obtained from HSC PDR3. To ensure reproducibility, the PREML dataset is archived on Zenodo at \url{https://doi.org/10.5281/zenodo.15426393}. Code and additional materials are available at: \url{https://github.com/projectultra/PREML}.

\bibliographystyle{unsrt}  
\bibliography{references}

\begin{thebibliography}{10}

\bibitem{LargeScaleStructure}
T.~M.~C. Abbott, M.~Aguena, S.~Allam, A.~Amon, F.~Andrade-Oliveira, J.~Asorey, S.~Avila, G.~M. Bernstein, E.~Bertin, A.~Brandao-Souza, D.~Brooks, D.~L. Burke, J.~Calcino, H.~Camacho, A.~Carnero~Rosell, D.~Carollo, M.~Carrasco~Kind, J.~Carretero, F.~J. Castander, R.~Cawthon, K.~C. Chan, A.~Choi, C.~Conselice, M.~Costanzi, M.~Crocce, L.~N. da~Costa, M.~E.~S. Pereira, T.~M. Davis, J.~De~Vicente, S.~Desai, H.~T. Diehl, P.~Doel, K.~Eckert, J.~Elvin-Poole, S.~Everett, A.~E. Evrard, X.~Fang, I.~Ferrero, A.~Fert\'e, B.~Flaugher, P.~Fosalba, J.~Garc\'{\i}a-Bellido, E.~Gaztanaga, D.~W. Gerdes, T.~Giannantonio, K.~Glazebrook, D.~Gomes, D.~Gruen, R.~A. Gruendl, J.~Gschwend, G.~Gutierrez, S.~R. Hinton, D.~L. Hollowood, K.~Honscheid, D.~Huterer, B.~Jain, D.~J. James, T.~Jeltema, N.~Kokron, E.~Krause, K.~Kuehn, O.~Lahav, G.~F. Lewis, C.~Lidman, M.~Lima, H.~Lin, M.~A.~G. Maia, U.~Malik, P.~Martini, P.~Melchior, J.~Mena-Fern\'andez, F.~Menanteau, R.~Miquel, J.~J. Mohr, R.~Morgan, J.~Muir, J.~Myles, A.~M\"oller, A.~Palmese,
  F.~Paz-Chinch\'on, W.~J. Percival, A.~Pieres, A.~A. Plazas~Malag\'on, A.~Porredon, J.~Prat, K.~Reil, M.~Rodriguez-Monroy, A.~K. Romer, A.~Roodman, R.~Rosenfeld, A.~J. Ross, E.~Sanchez, D.~Sanchez~Cid, V.~Scarpine, S.~Serrano, I.~Sevilla-Noarbe, E.~Sheldon, M.~Smith, M.~Soares-Santos, E.~Suchyta, M.~E.~C. Swanson, G.~Tarle, D.~Thomas, C.~To, M.~A. Troxel, B.~E. Tucker, D.~L. Tucker, I.~Tutusaus, S.~A. Uddin, T.~N. Varga, J.~Weller, and R.~D. Wilkinson.
\newblock Dark energy survey year 3 results: A 2.7
\newblock {\em Phys. Rev. D}, 105:043512, Feb 2022.

\bibitem{finkelstein2015evolution}
Steven~L Finkelstein, Russell~E Ryan, Casey Papovich, Mark Dickinson, Mimi Song, Rachel~S Somerville, Henry~C Ferguson, Brett Salmon, Mauro Giavalisco, Anton~M Koekemoer, et~al.
\newblock The evolution of the galaxy rest-frame ultraviolet luminosity function over the first two billion years.
\newblock {\em The Astrophysical Journal}, 810(1):71, 2015.

\bibitem{ivezic2019lsst}
{\v{Z}}eljko Ivezi{\'c}, Steven~M Kahn, J~Anthony Tyson, Bob Abel, Emily Acosta, Robyn Allsman, David Alonso, Yusra AlSayyad, Scott~F Anderson, John Andrew, et~al.
\newblock Lsst: from science drivers to reference design and anticipated data products.
\newblock {\em The Astrophysical Journal}, 873(2):111, 2019.

\bibitem{Euclid}
{Euclid Collaboration}, {Ilić, S.}, {Aghanim, N.}, {Baccigalupi, C.}, {Bermejo-Climent, J. R.}, {Fabbian, G.}, {Legrand, L.}, {Paoletti, D.}, {Ballardini, M.}, {Archidiacono, M.}, {Douspis, M.}, {Finelli, F.}, {Ganga, K.}, {Hernández-Monteagudo, C.}, {Lattanzi, M.}, {Marinucci, D.}, {Migliaccio, M.}, {Carbone, C.}, {Casas, S.}, {Martinelli, M.}, {Tutusaus, I.}, {Natoli, P.}, {Ntelis, P.}, {Pagano, L.}, {Wenzl, L.}, {Gruppuso, A.}, {Kitching, T.}, {Langer, M.}, {Mauri, N.}, {Patrizii, L.}, {Renzi, A.}, {Sirri, G.}, {Stanco, L.}, {Tenti, M.}, {Vielzeuf, P.}, {Lacasa, F.}, {Polenta, G.}, {Yankelevich, V.}, {Blanchard, A.}, {Sakr, Z.}, {Pourtsidou, A.}, {Camera, S.}, {Cardone, V. F.}, {Kilbinger, M.}, {Kunz, M.}, {Markovic, K.}, {Pettorino, V.}, {Sánchez, A. G.}, {Sapone, D.}, {Amara, A.}, {Auricchio, N.}, {Bender, R.}, {Bodendorf, C.}, {Bonino, D.}, {Branchini, E.}, {Brescia, M.}, {Brinchmann, J.}, {Capobianco, V.}, {Carretero, J.}, {Castander, F. J.}, {Castellano, M.}, {Cavuoti, S.}, {Cimatti, A.},
  {Cledassou, R.}, {Congedo, G.}, {Conselice, C. J.}, {Conversi, L.}, {Copin, Y.}, {Corcione, L.}, {Costille, A.}, {Cropper, M.}, {Da Silva, A.}, {Degaudenzi, H.}, {Dubath, F.}, {Duncan, C. A. J.}, {Dupac, X.}, {Dusini, S.}, {Ealet, A.}, {Farrens, S.}, {Fosalba, P.}, {Frailis, M.}, {Franceschi, E.}, {Franzetti, P.}, {Fumana, M.}, {Garilli, B.}, {Gillard, W.}, {Gillis, B.}, {Giocoli, C.}, {Grazian, A.}, {Grupp, F.}, {Guzzo, L.}, {Haugan, S. V. H.}, {Hoekstra, H.}, {Holmes, W.}, {Hormuth, F.}, {Hudelot, P.}, {Jahnke, K.}, {Kermiche, S.}, {Kiessling, A.}, {Kohley, R.}, {Kubik, B.}, {Kümmel, M.}, {Kurki-Suonio, H.}, {Laureijs, R.}, {Ligori, S.}, {Lilje, P. B.}, {Lloro, I.}, {Mansutti, O.}, {Marggraf, O.}, {Marulli, F.}, {Massey, R.}, {Maurogordato, S.}, {Meneghetti, M.}, {Merlin, E.}, {Meylan, G.}, {Moresco, M.}, {Morin, B.}, {Moscardini, L.}, {Munari, E.}, {Niemi, S. M.}, {Padilla, C.}, {Paltani, S.}, {Pasian, F.}, {Pedersen, K.}, {Percival, W.}, {Pires, S.}, {Poncet, M.}, {Popa, L.}, {Pozzetti, L.}, {Raison,
  F.}, {Rebolo, R.}, {Rhodes, J.}, {Roncarelli, M.}, {Rossetti, E.}, {Saglia, R.}, {Scaramella, R.}, {Schneider, P.}, {Secroun, A.}, {Seidel, G.}, {Serrano, S.}, {Sirignano, C.}, {Starck, J. L.}, {Tallada-Crespí, P.}, {Taylor, A. N.}, {Tereno, I.}, {Toledo-Moreo, R.}, {Torradeflot, F.}, {Valentijn, E. A.}, {Valenziano, L.}, {Verdoes Kleijn, G. A.}, {Wang, Y.}, {Welikala, N.}, {Weller, J.}, {Zamorani, G.}, {Zoubian, J.}, {Medinaceli, E.}, {Mei, S.}, {Rosset, C.}, {Sureau, F.}, {Vassallo, T.}, {Zacchei, A.}, {Andreon, S.}, {Balaguera-Antolínez, A.}, {Baldi, M.}, {Bardelli, S.}, {Biviano, A.}, {Borgani, S.}, {Bozzo, E.}, {Burigana, C.}, {Cabanac, R.}, {Cappi, A.}, {Carvalho, C. S.}, {Castignani, G.}, {Colodro-Conde, C.}, {Coupon, J.}, {Courtois, H. M.}, {Cuby, J.}, {de la Torre, S.}, {Di Ferdinando, D.}, {Dole, H.}, {Farina, M.}, {Ferreira, P. G.}, {Flose-Reimberg, P.}, {Galeotta, S.}, {Gozaliasl, G.}, {Graciá-Carpio, J.}, {Keihanen, E.}, {Kirkpatrick, C. C.}, {Lindholm, V.}, {Mainetti, G.}, {Maino, D.},
  {Martinet, N.}, {Maturi, M.}, {Metcalf, R. B.}, {Morgante, G.}, {Neissner, C.}, {Nightingale, J.}, {Nucita, A. A.}, {Potter, D.}, {Riccio, G.}, {Romelli, E.}, {Schirmer, M.}, {Schultheis, M.}, {Scottez, V.}, {Teyssier, R.}, {Tramacere, A.}, {Valiviita, J.}, {Viel, M.}, {Whittaker, L.}, and {Zucca, E.}
\newblock Euclid preparation - xv. forecasting cosmological constraints for the euclid and cmb joint analysis.
\newblock {\em A\&A}, 657:A91, 2022.

\bibitem{ivezic2018lsst1}
\v{Z}eljko Ivezi\'c and The LSST~Science Collaboration.
\newblock Lsst science requirements document lpm-17, 2018.
\newblock Accessed: 2025-05-04.

\bibitem{Pasquet2019}
J.~Pasquet, E.~Bertin, M.~Treyer, S.~Arnouts, and D.~Fouchez.
\newblock Photometric redshifts from sdss images using a convolutional neural network.
\newblock {\em Astronomy and Astrophysics}, 2019.

\bibitem{Schuldt2021}
S.~Schuldt et~al.
\newblock Photometric redshift estimation with a convolutional neural network: Netz.
\newblock {\em Astronomy and Astrophysics}, 2021.

\bibitem{Yao2023}
L.~Yao et~al.
\newblock Photometric redshift estimation of quasars with fused features from photometric data and images.
\newblock {\em Monthly Notices of the Royal Astronomical Society}, 2023.

\bibitem{AitOuahmed2023}
R.~Ait~Ouahmed, S.~Arnouts, J.~Pasquet, M.~Treyer, and E.~Bertin.
\newblock Multimodality for improved cnn photometric redshifts.
\newblock {\em Astronomy \& Astrophysics}, 2023.

\bibitem{Jones2024b}
E.~L. Jones et~al.
\newblock Improving photometric redshift estimation for cosmology with lsst using bayesian neural networks.
\newblock {\em The Astrophysical Journal}, 2024.

\bibitem{Jones2024a}
E.~Jones et~al.
\newblock Redshift prediction with images for cosmology using a bayesian convolutional neural network with conformal predictions.
\newblock {\em The Astrophysical Journal}, 974:159, 2024.

\bibitem{Zhou2022}
X.~Zhou et~al.
\newblock Photometric redshift estimates using bayesian neural networks in the csst survey.
\newblock {\em Research in Astronomy and Astrophysics}, 2022.

\bibitem{Luo2024}
Z.~Luo et~al.
\newblock Photometric redshift estimation for csst survey with lstm neural networks.
\newblock {\em Monthly Notices of the Royal Astronomical Society}, 535:1844--1855, 2024.

\bibitem{Dey2021}
B.~Dey et~al.
\newblock Photometric redshifts from sdss images with an interpretable deep capsule network.
\newblock {\em Monthly Notices of the Royal Astronomical Society}, 2021.

\bibitem{Hong2022}
S.~Hong et~al.
\newblock Photoredshift-mml: a multimodal machine learning method for estimating photometric redshifts of quasars.
\newblock {\em Monthly Notices of the Royal Astronomical Society}, 2022.

\bibitem{Henghes2022}
Henghes et~al.
\newblock Investigating deep learning methods for obtaining photometric redshift estimations from images.
\newblock {\em Monthly Notices of the Royal Astronomical Society}, 2022.

\bibitem{Hoyle2016}
B.~Hoyle.
\newblock Measuring photometric redshifts using galaxy images and deep st neural networks.
\newblock {\em Astronomy and Computing}, 2016.

\bibitem{DIsanto2017}
A.~D’Isanto and K.~Polsterer.
\newblock Photometric redshift estimation via deep learning.
\newblock {\em Astronomy and Astrophysics}, 2017.

\bibitem{aihara2022third}
Hiroaki Aihara, Yusra AlSayyad, Makoto Ando, Robert Armstrong, James Bosch, Eiichi Egami, Hisanori Furusawa, Junko Furusawa, Sumiko Harasawa, Yuichi Harikane, et~al.
\newblock Third data release of the hyper suprime-cam subaru strategic program.
\newblock {\em Publications of the Astronomical Society of Japan}, 74(2):247--272, 2022.

\bibitem{do2024galaxiesmldatasetgalaxyimages}
Tuan Do, Bernie Boscoe, Evan Jones, Yun~Qi Li, and Kevin Alfaro.
\newblock Galaxiesml: a dataset of galaxy images, photometry, redshifts, and structural parameters for machine learning, 2024.

\end{thebibliography}

\end{document}